\documentstyle[twoside,fleqn,espcrc2,psfig]{article}
\newcommand{\bra}{\langle}
\newcommand{\ket}{\rangle}
\newcommand{\half}{\frac{1}{2}}

\newcommand{\cl}{{\rm cl}}
\newcommand{\rmd}{{\rm d}}       
\newcommand{\be}{\begin{equation}}
\newcommand{\ee}{\end{equation}}
\newcommand{\bea}{\begin{eqnarray}}
\newcommand{\eea}{\end{eqnarray}}
\newcommand{\bean}{\begin{eqnarray*}}
\newcommand{\eean}{\end{eqnarray*}}

\newcommand{\vecp}{{\mathbf p}}

\newcommand{\vecx}{{\mathbf x}}

\newcommand{\vecnul}{{\mathbf 0}}
\newcommand{\om}{\omega}

\title{\vskip-5mm 
Spectral function at high temperature\thanks{
Presented at Lattice 2001, 19-24 August 2001, Berlin.
}\vskip-2.3cm\hfill\small HD-THEP-01-39\vskip2.3cm
}
\author{
Gert Aarts\address{Institut f\"ur theoretische Physik, Universit\"at 
Heidelberg,\\
Philosophenweg 16, 69120 Heidelberg, Germany}
}

\begin{document}

\begin{abstract}
For a weakly coupled quantum field at high temperature the classical
approximation offers a possibility to gain insight into nonperturbative
real-time dynamics.  I use this to present a nonperturbative approach
to the computation of spectral functions in real time.  Results are
shown for a scalar field in $2+1$ dimensions.
\end{abstract}
\maketitle

1.\ 
In thermal equilibrium spectral functions contain all available
information since other real- (and imaginary-) time correlators are
related to them via the KMS condition.  A nonperturbative computation of
spectral functions seems therefore desirable; nonperturbative in order 
to:
\begin{itemize}
\item[i)] deal with a truly nonperturbative theory, such as
the infrared sector of high-temperature nonabelian gauge fields,
\item[ii)] investigate the utility of (resummed) perturbative ideas,
since weak coupling $+$ high temperature $\neq$ simple. 
\end{itemize}
For the calculation of static quantities, such as free energies, phase
diagrams and screening masses, the euclidean lattice formulation of
finite-temperature quantum fields offers a first-principle approach. For
intrinsically {\em real-time} correlators, such as the spectral function,
no first-principle nonperturbative formulation exists. In the last ${\cal
O}(10)$ years the classical approximation at high temperature and weak
coupling has provided much insight into real-time dynamics, see
\cite{Bodeker:2001pa,Yaffe:2001} for recent reviews. In this talk I
discuss how spectral functions can be computed in the classical
approximation and what can be learned from them. Details can be found in
\cite{Aarts:2001yx}.

2.\ 
The spectral function for a bosonic operator $O$ can be defined as the
thermal expectation value
\be
\rho(x-y) = i\bra [O(x),O^\dagger(y)]_-\ket.
\ee
Other two-point functions can be expressed in terms of it using the KMS
condition, leading to the appearance of the Bose distribution function
$n(\om) = 1/[e^{\om/T}-1]$. For example, the euclidean correlator
$D(\tau, \vecx) = \bra O(\tau,\vecx) O^\dagger(0,\vecnul)\ket_E$, with
imaginary time $\tau\in [0,1/T]$, is related to the spectral function 
via the integral equation
\be
D(\tau, \vecp) = \int_0^\infty \frac{\rmd \om}{2\pi i}\,
K(\tau,\om)\rho(\om,\vecp),
\ee
with the kernel 
\be
K(\tau,\om) = n(\om)\left[e^{\om\tau} +
e^{-\om(\tau-1/T)}\right]
\ee
obeying $K(\tau, \om) = -K(\tau, -\om) = K(1/T-\tau, \om)$.
The extraction of the spectral function from the euclidean
lattice correlator is a highly nontrivial inversion problem,
currently tackled  using the Maximal Entropy Method
\cite{MEM}. 
In real time the spectral function would be accessible directly without
the need to solve an integral equation. 

The classical approximation provides a nonperturbative computational
scheme in real time. In the classical limit $-i$ times the commutator
is replaced with the Poisson bracket and the classical spectral function
reads
\be
\label{eqPB}
\rho_\cl(x-y) = -\bra \{O(x),O^\dagger(y)\}\ket_\cl.
\ee
The time evolution is determined by the classical equations of
motion, evolving from some initial condition. 
The brackets $\bra\cdot\ket_\cl$ denote a Boltzmann-weighted average over
these initial conditions. 

In perturbation theory the relation between classical and quantum physics
at finite temperature can be seen clearly. Consider a 
self-interacting scalar field with coupling constant $\lambda$. The
effective expansion parameter is $\hbar\lambda n(\hbar\om)$ 
(with the $\hbar$ dependence indicated explicitly). As a result, in the
classical limit each loop contributes with a proportionality factor 
$\lambda T/\om$. Note that this implies that the
infrared sector ($\hbar\om/T \ll 1$) is essentially classical but that  
in the ultraviolet problems are encountered 
(the Rayleigh-Jeans divergence, see below).

\begin{figure}
\centerline{\psfig{figure=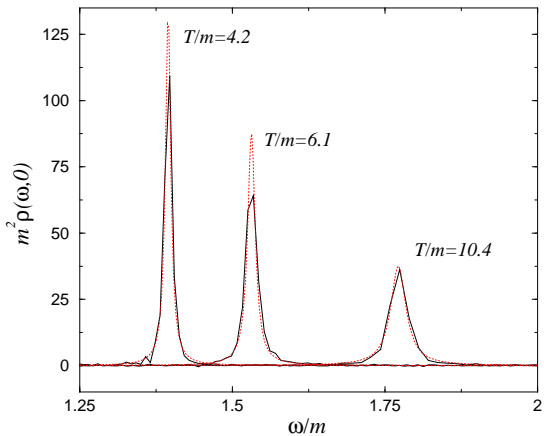,height=5.9cm}}
\vspace{-1cm}
\caption{
Spectral functions $\rho_\cl(\om,\vecnul)$ for various temperatures
$T$. Fits to a Breit-Wigner function are shown with dotted lines. 
}
\label{fig3}
\end{figure}

For actual calculations, the Poisson bracket in Eq.\ (4) appears difficult
to use numerically, but in equilibrium the KMS condition comes to the 
rescue. In the classical limit the KMS relation relates the statistical
correlator 
$
S(x-y) = \bra O(x)O^\dagger(y)\ket_\cl 
$
and the spectral function in momentum space as
\be
-i\frac{T}{\om}\rho_\cl(\om,\vecp) = S(\om,\vecp).
\ee
or equivalently in real-space as
\be
\rho_\cl(t,\vecx) = -\frac{1}{T}\partial_t S(t,\vecx).
\ee
The latter relation, valid for arbitrary $O$, allows an easy
calculation of classical spectral functions. 

\begin{figure} 
\centerline{\psfig{figure=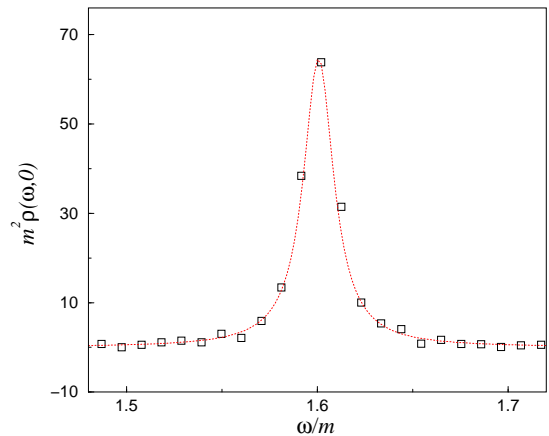,height=5.9cm}} 
\vspace{-1cm}
\caption{As in Fig.\ 1 at $T/m=7.2$. 
The actual data points are presented with squares.
} 
\label{fig2} 
\end{figure}

3.\ 
To illustrate the method we consider a simple real scalar field with a
$\lambda\phi^4$/4!-interaction in the symmetric phase in $2+1$
dimensions (note that in a classical theory the dimensionless coupling 
$\lambda/m$ can be taken equal to 1 without loss of generality).
We focus on the one-particle spectral function. According to the
classical KMS condition we have to compute
\bea
\nonumber
\rho_\cl(t,\vecx)\!\!\!\! &=& \!\!\!\!-\bra \{\phi(t,\vecx),
\phi(0,\vecnul)\}\ket_\cl
\\
\!\!\!\!&=&\!\!\!\!
-\frac{1}{T}\bra \pi(t,\vecx) \phi(0,\vecnul)\ket_\cl,
\eea
where $\pi=\partial_t\phi$ is the canonical momentum. The right-hand-side
is a simple correlation function that can be computed numerically
without any problems \cite{Aarts:2001yx}. The theory is
defined on a spatial lattice of $N\times N$ sites and lattice spacing
$a$ (we used $N=128, ma=0.2$, and periodic boundary conditions). The
classical equations of motion are solved with a leapfrog algorithm
with time step $a_0/a=0.1$. Thermal initial configurations are
generated with the Kramers equation algorithm (the plots shown are
obtained with 2000 independently thermalized initial configurations). On
the
lattice the spectral function (7) is symmetrized:
\bea
\nonumber
\rho_{\cl,\rm lat}(t,\vecx) \!\!\!\!&=& \\
&& 
\!\!\!\!\!\!\!\!\!\!\!\!
\!\!\!\!\!\!\!\!\!\!\!\!
\!\!\!\!\!\!\!\!\!\!
 -\frac{1}{T}\left\bra
\pi(t+\half a_0,\vecx) \half\left[\phi(0,\vecnul)+ \phi(a_0,\vecnul)\right]
\right\ket_\cl\!.
\eea
The simulations give the spectral function as a correlation function
in real time. A sine transform (using symmetry properties
of the spectral function under time reflection and complex conjugation)
yields the desired result as a function of frequency.  In Fig.~1 the
spectral function at zero momentum is shown for various temperatures and
in Fig.~2 a magnification with the actual data points indicated is
presented. The maximal (real) time $t_{\rm max}$ used in the analysis
constrains the resolution in frequency space to $\Delta\om=\pi/t_{\rm
max}$, but this poses no problem ($\Delta \om/m\approx 0.01$ in Fig.~2).

\begin{figure}
\centerline{\psfig{figure=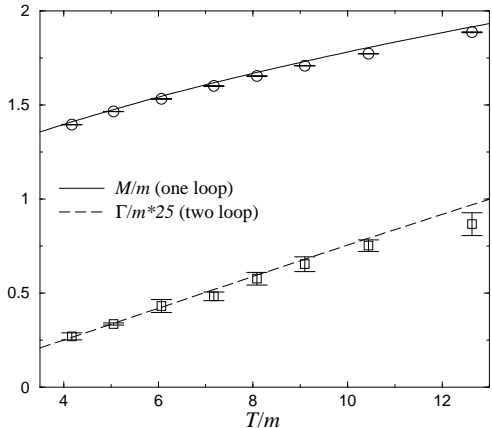,height=5.9cm}}
\vspace{-1cm}
\caption{
Temperature dependence of the mass $M$ (circles) and width $\Gamma$
(squares, multiplied by 25 for clarity), obtained from fits to a
Breit-Wigner function. Errors are determined with a jackknife analysis. 
The resummed perturbative predictions are shown with lines.
}
\label{fig4}
\end{figure}                               

4.\ 
It is now clear that spectral functions can be computed 
nonperturbatively in classical thermal field theory. The question is what 
we may learn for hot {\em quantum} fields. 

As alluded to above, the setup of perturbation theory in the quantum and 
classical case is similar and the results may be used to address the
applicability of (resummed) perturbation theory. In the weak-coupling 
regime the spectral function is dominated by the
plasmon. A simple way to parametrize it (ignoring the proper
analytical structure) is by a Breit-Wigner spectral function. At zero
momentum it reads 
\be
\rho_{\rm BW}(\om) =
\frac{2\om\Gamma}{(\om^2-M^2)^2+\om^2\Gamma^2}.
\ee
Fits of the data to a Breit-Wigner function are shown in Figs.\ 1,2,
and yield nonperturbatively determined values of the plasmon mass and
width. These can be compared with the perturbative predictions.  
At lowest order the mass parameter $M$ is determined from the classical
limit of the standard one-loop gap equation. 
The leading contribution to the width $\Gamma$ comes from the two-loop
setting-sun diagram and on dimensional grounds it is $\Gamma =
c\lambda^2T^2/M^3$, with a small coefficient $c=(3-2\sqrt{2})/(32\pi)$
\cite{Aarts:2001yx}.
A comparison between the perturbatively and nonperturbatively determined
values of the effective mass and width is shown in Fig.\ 3. A nice
agreement can be seen, indicating that for this range of parameters
perturbation theory is reliable.

5.\ 
It is straightforward to extend the calculation to more complicated
spectral functions in scalar and gauge theories (for lattice fermions in 
real time, see \cite{Aarts:1999td}).
Transport coefficients can be defined from the zero-momentum and
zero-frequency limit of equilibrium spectral functions of appropriate
composite operators. Simple power counting shows that these quantities are
dominated by hard ($\sim T$) momenta. In a classical calculation they will
therefore be sensitive to the lattice regulator. 
In fact, such dependence has already been encountered since the
effective mass parameter $M$ depends (logarithmically in $2+1$
dimensions) on the lattice cutoff. However, as can be seen from Fig.\ 3,
this does not automatically imply that the classical findings cannot be
used: analytical perturbative and numerical nonperturbative calculations
can  still be compared, provided the role of the lattice regulator is
incorporated properly.

\vspace{0.3cm}
\noindent {\bf Acknowledgements}: 
%It is a pleasure to thank Nucu Stamatescu for discussions.  This work 
%was supported by 
%the TMR network {\em Finite Temperature Phase Transitions in Particle 
%Physics}, 
%EU contract no.\ FMRX-CT97-0122.
I thank Nucu Stamatescu for discussions.  
Supported by the TMR network, 
%{\em Finite Temperature Phase Transitions in Particle 
%Physics}, 
EU contract no.\ FMRX-CT97-0122.


\begin{thebibliography}{10}

%\cite{Bodeker:2001pa}
\bibitem{Bodeker:2001pa}
D.~B\"odeker,
%``Non-equilibrium field theory,''
Nucl.\ Phys.\ Proc.\ Suppl.\  {\bf 94} (2001) 61
[hep-lat/0011077].
%%CITATION = HEP-LAT 0011077;%%

%\cite{Yaffe:2001}
\bibitem{Yaffe:2001}
L.~G.~Yaffe, these Proceedings. 

%\cite{Aarts:2001yx}
\bibitem{Aarts:2001yx}
G.~Aarts,
Phys.\ Lett.\ B {\bf 518} (2001) 315
[hep-ph/0108125].
%%CITATION = HEP-PH 0108125;%%      

\bibitem{MEM}
%\cite{Wetzorke:2000ez}
%\bibitem{Wetzorke:2000ez}
I.~Wetzorke and F.~Karsch,
%``Testing MEM with diquark and thermal meson correlation functions,''
hep-lat/0008008,
%%CITATION = HEP-LAT 0008008;%%
and various contributions to these Procs.

%\cite{Aarts:1999td}
\bibitem{Aarts:1999td}
G.~Aarts and J.~Smit,
%``Real-time dynamics with fermions on a lattice,''
Nucl.\ Phys.\ B {\bf 555} (1999) 355
[hep-ph/9812413].
%%CITATION = HEP-PH 9812413;%%




\end{thebibliography}
\end{document}